\documentclass{Interspeech}

\interspeechcameraready 
\usepackage{cite}
\usepackage{multirow, multicol}
\usepackage{makecell}
\usepackage{bbding}
\usepackage{enumitem}
\PassOptionsToPackage{hyphens}{url}
\usepackage{url}
\usepackage{hyperref}

\title{Improving Noise Robustness of LLM-based Zero-shot TTS via Discrete Acoustic Token Denoising\thanks{$^*$ Corresponding author.}}

\author{Ye-Xin}{Lu}
\author{Hui-Peng}{Du}
\author{Fei}{Liu}
\author{Yang}{Ai$^*$}
\author{Zhen-Hua}{Ling}

\affiliation{National Engineering Research Center of Speech and Language Information Processing}{\\University of Science and Technology of China}{Hefei, P. R. China}
\email{\{yxlu0102, redmist, fliu215\}@mail.ustc.edu.cn, \{yangai, zhling\}@ustc.edu.cn}

\begin{document}
\maketitle
\begin{abstract}
Large language model (LLM) based zero-shot text-to-speech (TTS) methods tend to preserve the acoustic environment of the audio prompt, leading to degradation in synthesized speech quality when the audio prompt contains noise. In this paper, we propose a novel neural codec-based speech denoiser and integrate it with the advanced LLM-based TTS model, LauraTTS, to achieve noise-robust zero-shot TTS. The proposed codec denoiser consists of an audio codec, a token denoiser, and an embedding refiner. The token denoiser predicts the first two groups of clean acoustic tokens from the noisy ones, which can serve as the acoustic prompt for LauraTTS to synthesize high-quality personalized speech or be converted to clean speech waveforms through the embedding refiner and codec decoder. Experimental results show that our proposed codec denoiser outperforms state-of-the-art speech enhancement (SE) methods, and the proposed noise-robust LauraTTS surpasses the approach using additional SE models.

\end{abstract}

\section{Introduction}
Zero-shot text-to-speech (TTS) synthesis \cite{casanova2021sc, casanova2022yourtts, wang2023neural, du2023lauragpt, ju2024naturalspeech} aims to synthesize any speaker's voice based on few seconds of audio prompt of the speaker.
Typically trained on extensive high-quality speech data, these models face challenges in real-life scenarios where obtaining clean audio prompts is challenging. 
Recorded audio prompts are often distorted by intrusive noises, leading to a degradation in the synthesized speech quality.
Consequently, some noise-robust zero-shot TTS methods \cite{fujita2024noise, pankov2024dino} have been proposed, aiming to synthesize high-quality personalized speech from noisy audio prompts.
These methods enhance the noise robustness of zero-shot TTS systems by training a noise-robust speaker encoder to extract robust speaker embeddings from noisy audio prompts. 
Although having shown promising performance, this category of approaches is only applicable to zero-shot TTS systems that are based on speaker embeddings.

\begin{figure}[t!]
  \centering
  \includegraphics[width=\linewidth]{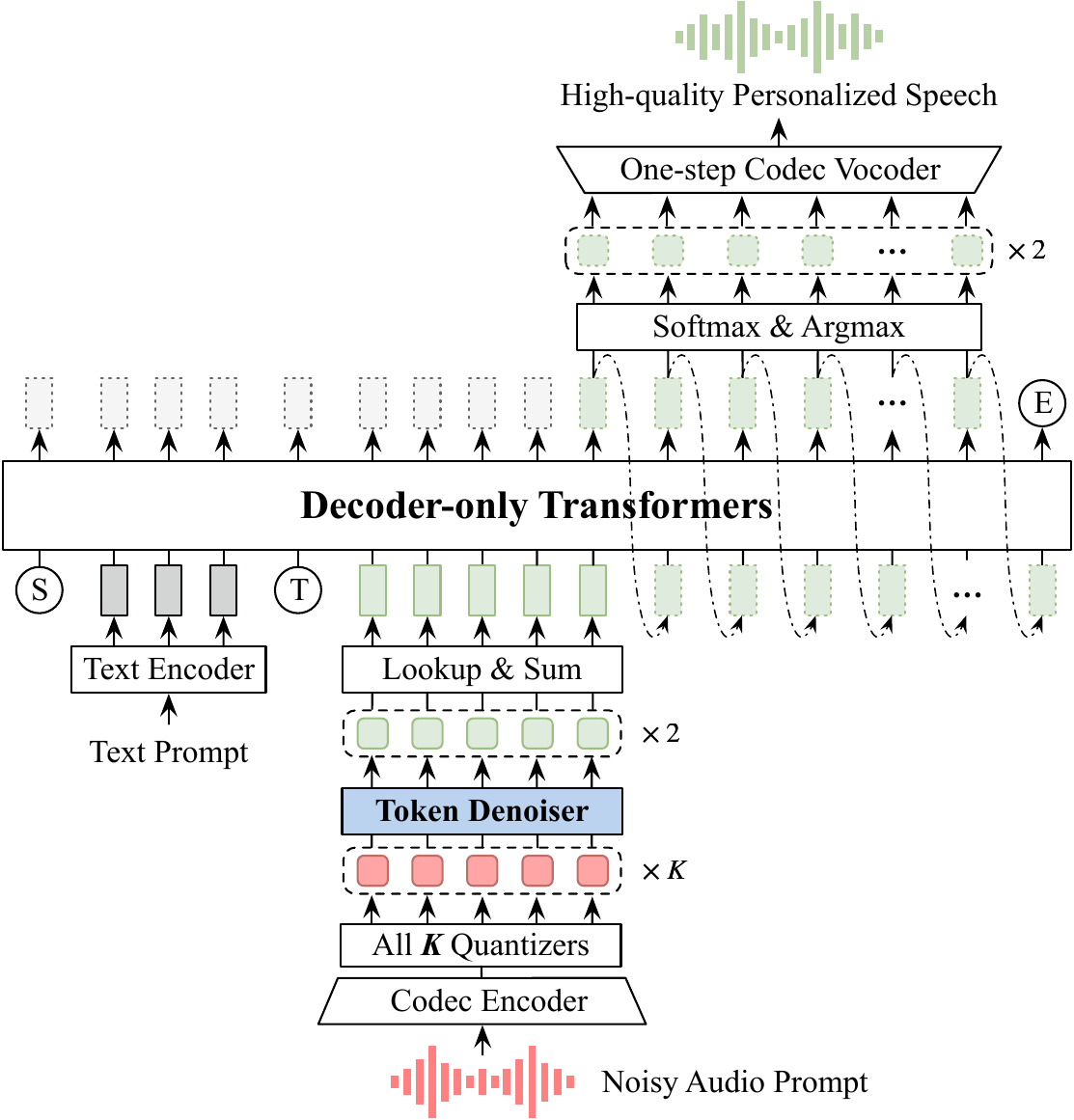}
  \caption{Noise-robust zero-shot TTS synthesis process of the proposed NR-LauraTTS, where $\textcircled{S}$, $\textcircled{T}$, and $\textcircled{E}$ denote the ``start of sequence'', ``turn of speech'', and ``end of sequence'' tokens.} 
  \label{fig: nr-lauratts}
\end{figure}

\begin{figure*}[ht!]
  \centering
  \includegraphics[width=0.9\textwidth]{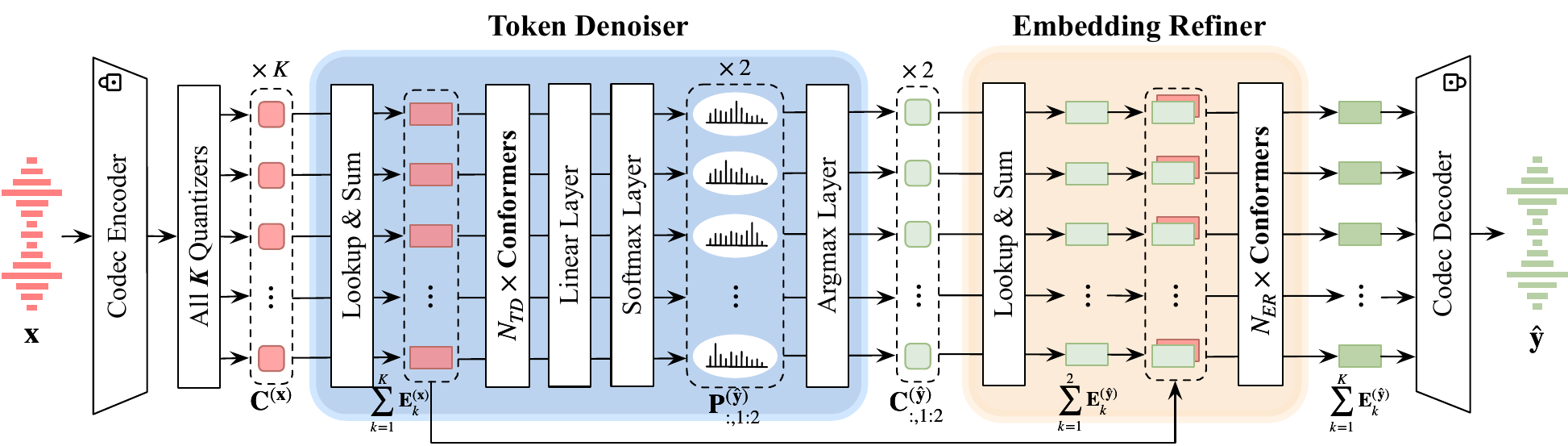}
  \caption{Overall structure of the proposed codec denoiser.} 
  \label{fig: codec-denoiser}
\end{figure*}

Recently, large language model (LLM) based TTS methods \cite{wang2023neural, du2023lauragpt, ju2024naturalspeech} have emerged as the mainstream approach for zero-shot TTS due to their exceptional naturalness and speaker similarity in synthesized speech.
They treat TTS as a conditional codec language modeling task, employing discrete acoustic tokens extracted by neural codecs \cite{zeghidour2021soundstream, defossezhigh} as the intermediate representation. 
During generation, this category of methods demonstrates a strong in-context learning capability and tends to preserve the acoustic environment of the audio prompt, making them more noise-sensitive than speaker-embedding-based approaches.
Research on how to improve the noise robustness of LLM-based zero-shot TTS is still limited.
A common approach involves employing a speech enhancement (SE) model to denoise the noisy audio prompt before feeding it to the TTS models \cite{botinhao2016speech, valentini2016investigating, valentini2018speech}.
However, even the state-of-the-art (SOTA) SE models \cite{cao2022cmgan, lu2023mp, lu2023explicit} would introduce artifacts \cite{iwamoto2022bad, sato2022learning} to the enhanced audio prompt and harm the TTS performance.
Moreover, enhancing noisy audio prompts at the signal level would introduce significant computational complexity, impacting the overall efficiency of the noise-robust TTS pipeline.

In this paper, we propose to enhance the noise robustness of LLM-based zero-shot TTS at the level of discrete acoustic tokens.
Specifically, we utilize the advanced LLM-based zero-shot TTS system, LauraTTS \cite{du2023lauragpt}, as the TTS backbone.
We first propose a neural codec-based speech denoiser, dubbed codec denoiser, which comprises an audio codec, a token denoiser, and an embedding refiner, achieving speech enhancement (SE) at the discrete acoustic token domain.
The token denoiser predicts the first two groups of enhanced acoustic tokens from all groups of noisy ones, which are then processed by the embedding refiner to predict a summed enhanced acoustic embedding. 
Finally, the codec decoder reconstructs the enhanced speech waveform.
Additionally, we integrate the codec denoiser into LauraTTS to construct a noise-robust zero-shot TTS system, NR-LauraTTS, which is capable of synthesizing high-quality personalized speech from noisy audio prompts.
Experimental results demonstrate that our proposed codec denoiser surpasses SOTA SE methods in enhanced speech quality and model's computational complexity, and the NR-LauraTTS excels in synthesized speech naturalness, intelligibility, and speaker similarity compared to the approach using additional SE models, with only a minimal complexity increment.

\vspace{-1mm}
\section{Methodology}
In this section, we first elaborate on the noise-robust zero-shot TTS synthesis process of the NR-LauraTTS model, and then describe in detail the model structure and training criteria of the proposed codec denoiser.

\subsection{Noise-Robust LauraTTS}
The overall noise-robust zero-shot TTS synthesis process of the proposed NR-LauraTTS is depicted in Fig.~\ref{fig: nr-lauratts}, which comprises a pre-trained LauraTTS and a token denoiser trained within the codec denoiser.
The LauraTTS is based on the Encodec model \cite{defossezhigh}, which employs residual vector quantization (RVQ) with $K$ vector quantizers (VQs), each having a codebook size of $C$, to extract discrete acoustic tokens.
Due to the RVQ in Encodec, quantized tokens from initial quantizers capture essential acoustic properties such as speaker identity, while the subsequent quantizers learn fine acoustic details.
Therefore, LauraTTS only predicts the first two groups of target acoustic tokens, then uses a one-step codec vocoder to estimate the summed acoustic embedding of all target acoustic token groups, and finally reconstructs the speech waveform.

For synthesizing high-quality personalized speech from noisy audio prompts, the codec encoder with $K$ VQs first encodes the noisy audio prompt into $K$ groups of noisy acoustic tokens, which are then enhanced by the token denoiser to produce the first two groups of enhanced acoustic tokens.
LauraTTS takes the text prompt and two groups of enhanced acoustic tokens as inputs, auto-regressively predicting the first two groups of target acoustic tokens.
Finally, these target acoustic tokens are processed by the one-step codec vocoder, with the condition of the text prompt, to reconstruct the high-quality personalized speech waveform. 

\subsection{Codec denoiser}
The overall structure of the proposed codec denoiser is illustrated in Fig.~\ref{fig: codec-denoiser}.
The codec denoiser aims to enhance the noisy speech signal $\mathbf{x} \in \mathbb{R}^{L}$ and recover the clean speech signal $\mathbf{y} \in \mathbb{R}^{L}$, where $L$ represents the waveform sequence length.
Although several methods have been proposed for SE in the discrete acoustic token domain \cite{li2024masksr, yang2024genhancer}, they typically predict tokens from all VQs. 
As the number of VQs grows, the difficulty of token prediction rises and the computational complexity of the model climbs.
Consequently, to overcome the prediction challenge posed by the multi-modal distribution of codec tokens and reduce the model complexity, inspired by LauraTTS \cite{du2023lauragpt}, we design a token denoiser to only predict the first two groups of clean acoustic tokens.
An embedding refiner is then employed to refine acoustic details, ultimately restoring the clean speech waveforms through the codec decoder.

Specifically, we first utilize the codec encoder with $K$ VQs to encode the noisy speech signal $\mathbf{x}$ and the clean speech signal $\mathbf{y}$ into discrete acoustic tokens $\mathbf{C}^{(\mathbf{x})} \in \mathbb{Z}^{T \times K}$ and $\mathbf{C}^{(\mathbf{y})} \in \mathbb{Z}^{T \times K}$, where their elements $c^{(\mathbf{x})}_{t,k}, c^{(\mathbf{y})}_{t,k} \in \{1, 2, ..., C\}$, with $1 \leq t \leq T, 1 \leq k \leq K$.
Here, $T = L/M$ represents the downsampled sequence length and $M$ represents the codec downsampling rate.
Subsequently, the token denoiser takes all $K$ groups of noisy acoustic tokens $\mathbf{C}^{(\mathbf{x})}$ as input and predicts the first two groups of enhanced acoustic tokens $\mathbf{C}^{(\mathbf{\hat{y}})}_{:, 1:2} \in \mathbb{Z}^{T \times 2}$, where $\mathbf{\hat{y}} \in \mathbb{R}^{L}$ denotes the enhanced speech signal.
The embedding refiner then predicts the summed embedding of all $K$ groups of enhanced acoustic tokens $\sum_{k=1}^{K}\mathbf{E}^{(\mathbf{\hat{y}})}_{k} \in \mathbb{R}^{T \times D}$, where $\mathbf{E}^{(\mathbf{\hat{y}})}_{k} \in \mathbb{R}^{T \times D}$ denotes the acoustic embedding corresponding to the enhanced acoustic token of the $k$-th VQ $\mathbf{c}^{(\mathbf{\hat{y}})}_{:, k}$, and $D$ represents the code vector dimension.
Finally, $\sum_{k=1}^{K}\mathbf{E}^{(\mathbf{\hat{y}})}_{k}$ is reconstructed to $\mathbf{\hat{y}}$ using the codec decoder. 
The details of the token denoiser and embedding refiner, and the training criteria of the codec denoiser are described as follows.

\subsubsection{Token Denoiser}
The token denoiser uses the Conformer \cite{gulati2020conformer} blocks as the backbone to predict the first two groups of enhanced acoustic tokens $\mathbf{C}^{(\mathbf{\hat{y}})}_{:, 1:2}$ from all groups of noisy acoustic tokens $\mathbf{C}^{(\mathbf{x})}$.
It first uses the pre-trained $K$ codebooks to lookup $K$ noisy acoustic embeddings and then adds them up to obtain a summed noisy acoustic embedding $\sum_{k=1}^{K}\mathbf{E}^{(\mathbf{x})}_{k} \in \mathbb{R}^{T \times D}$, which then undergoes $N_{TD}$ Conformer blocks, a linear layer, and a Softmax layer to obtain the probability distribution of the first two groups of the enhanced acoustic tokens $\mathbf{P}^{(\mathbf{\hat{y}})}_{:, 1:2} \in \mathbb{R}^{T \times 2 \times C}$, with the value of its element vector $\mathbf{p}^{(\mathbf{\hat{y}})}_{t, k} \in \mathbb{R}^{C}$ range from 0 to 1, where $1 \leq t \leq T, 1 \leq k \leq 2$.
Finally, $\mathbf{P}^{(\mathbf{\hat{y}})}_{:, 1:2}$ is converted to $\mathbf{C}^{(\mathbf{\hat{y}})}_{:, 1:2}$ through an Argmax layer, where the index of the highest probability in $\mathbf{p}^{(\mathbf{\hat{y}})}_{t, k}$ is selected as the predicted token.

\subsubsection{Embedding Refiner}
The embedding refiner also uses the Conformer blocks as the backbone.
It takes the first two groups of enhanced acoustic tokens $\mathbf{C}^{(\mathbf{\hat{y}})}_{:, 1:2}$ as input, initially utilizes the first two codebooks to lookup the corresponding enhanced acoustic embeddings and obtain the summed embedding $\sum_{k=1}^{2}\mathbf{E}^{(\mathbf{\hat{y}})}_{k}$.
To fully utilize the information in the noisy acoustic tokens $\mathbf{C}^{(\mathbf{x})}$, we concatenate $\sum_{k=1}^{2}\mathbf{E}^{(\mathbf{\hat{y}})}_{k}$ with the summed noisy embedding $\sum_{k=1}^{K}\mathbf{E}^{(\mathbf{x})}_{k}$ as the input conditions, which undergo $N_{ER}$ Conformer blocks to predict the summed enhanced embedding $\sum_{k=1}^{K}\mathbf{E}^{(\mathbf{\hat{y}})}_{k}$.

\subsubsection{Training Criteria}
For training the codec denoiser, we employ a pre-trained Encodec and use the teacher-forcing strategy to jointly train the token denoiser and the embedding refiner by randomly replacing the predicted acoustic tokens $\mathbf{C}^{(\mathbf{\hat{y}})}_{:, 1:2}$ with the corresponding clean ones $\mathbf{C}^{(\mathbf{y})}_{:, 1:2}$.
We first define the cross-entropy (CE) loss between the predicted probability distribution of the first two groups of enhanced acoustic tokens $\mathbf{P}^{(\mathbf{\hat{y}})}_{:, 1:2}$ and the corresponding target probability distribution $\mathbf{P}^{(\mathbf{y})}_{:, 1:2}$ as:
\begin{equation}
    \mathcal{L}_{CE} = \sum^{2}_{k=1} \sum^{T}_{t=1} CrossEntropy(\mathbf{p}^{(\mathbf{\hat{y}})}_{t, k}, \mathbf{p}^{(\mathbf{y})}_{t, k}),
\end{equation}
where $\mathbf{P}^{(\mathbf{y})}_{:, 1:2}$ is obtained by applying one-hot encoding on the first two groups of the clean acoustic tokens $\mathbf{C}^{(\mathbf{y})}_{:, 1:2}$.
Subsequently, we define the embedding refinement (ER) loss between the summed enhanced acoustic embedding $\sum_{k=1}^{K}\mathbf{E}^{(\mathbf{\hat{y}})}_{k}$ and the summed clean acoustic embedding $\sum_{k=1}^{K}\mathbf{E}^{(\mathbf{y})}_{k}$ as:
\begin{equation}
\mathcal{L}_{ER} = \bigg\Vert \sum_{k=1}^{K} (\mathbf{E}^{(\mathbf{\hat{y}})}_{k} - \mathbf{E}^{(\mathbf{y})}_{k}) \bigg\Vert_1 + \bigg\Vert \sum_{k=1}^{K} (\mathbf{E}^{(\mathbf{\hat{y}})}_{k} - \mathbf{E}^{(\mathbf{y})}_{k}) \bigg\Vert_\mathrm{F},
\end{equation}
where $\Vert \cdot \Vert_{\mathrm{F}}$ represents the Frobenius norm.
The final loss for training the codec denoiser is defined as the weighted sum of $\mathcal{L}_{CE}$ and $\mathcal{L}_{ER}$ as:
\begin{equation}
    \mathcal{L} = \lambda_1\mathcal{L}_{CE} + \lambda_2\mathcal{L}_{ER},
\end{equation}
where $\lambda_1, \lambda_2$ are hyper-parameters.

\section{Experiments}
\subsection{Dataset and Experimental Setup}
Due to the high training cost of LLM-based TTS models, we pre-trained LauraTTS using the LibriLight dataset \cite{kahn2020libri}, which contains about 60,000 hours of English speech data from over 7,000 speakers, and constructed a noisy-clean dataset based on the LibriTTS-R dataset \cite{koizumi2023libritts} to train the codec denoiser.
For training and validation, we utilized the 580-hour training set of the 15-hour development set of LibriTTS-R and incorporated noise data from the Deep Noise Suppression (DNS) Challenge 2022 dataset \cite{dubey2022icassp} to create a noisy training set and validation set, with the signal-to-noise ratio (SNR) uniformly distributed between -5 dB and 15 dB.  
For testing, we used the clean data from the ``test-clean'' set and incorporated noise from the WHAM! dataset \cite{wichern2019wham} to build a noisy test set with SNR uniformly distributed between 0 dB and 20 dB.
All the speech data was resampled to 16 kHz.

To extract discrete acoustic tokens from the speech signals, we utilized the pre-trained Encodec model from the official implementation of Funcodec \footnote{\href{https://github.com/modelscope/FunCodec}{https://github.com/modelscope/FunCodec}.} \cite{du2024funcodec}.
The Encodec achieved high-fidelity 16 kHz speech encoding with a codec downsampling rate $M$ of 640, a VQ count $K$ of 32, a codebook size $C$ of 1024, and a codec embedding dimension $D$ of 128.
Using the pre-trained Encodec to extract discrete acoustic tokens, we first trained the LauraTTS model for 80 epochs on the LibriLight dataset. 
Subsequently, we trained the proposed codec denoiser on the constructed noisy-clean dataset, with the number of Conformer blocks $N_{TD}$ set to 12 and $N_{CV}$ set to 6.
The hyper-parameters in the final loss, $\lambda_1$ and $\lambda_2$, were set to 1.0 and 0.5, respectively.
The codec denoiser model was trained for 60 epochs using the Adam optimizer, with a peak learning rate of 0.001 and 10,000 warm-up steps. 
\footnote{Audio samples of the proposed NR-LauraTTS can be accessed at \href{https://yxlu-0102.github.io/NR-LauraTTS}{https://yxlu-0102.github.io/NR-LauraTTS}.}

\begin{table}[t!]\scriptsize
  \caption{Objective evaluation results for the SE task.}
  \label{tab: se}
  \centering
  \resizebox{0.9\linewidth}{!}{
  \begin{tabular}{c|ccccc}
    \toprule
    Method & SIG  & BAK  & OVRL & FLOPs \\
    \midrule
    Noisy  & 3.01 & 2.08 & 2.05 & -     \\
    \midrule
    CMGAN \cite{cao2022cmgan}  & 3.54 & 4.05 & 3.26 & 31.68 G \\
    MP-SENet \cite{lu2023mp}   & 3.56 & 4.09 & 3.30 & 38.93 G \\
    \textbf{Codec Denoiser}    & \textbf{3.62} & \textbf{4.11} & \textbf{3.36} & \textbf{9.96 G} \\
    \bottomrule
  \end{tabular}}
\end{table}

\begin{table}[t!]\small
  \caption{Objective evaluation results for the proposed codec denoiser with different groups of predicted tokens.}
  \label{tab: ablation}
  \centering
  \resizebox{\linewidth}{!}{
  \begin{tabular}{c|ccccc}
    \toprule
    \# Predicted Token Groups  & SIG & BAK & OVRL & FLOPs \\
    \midrule
    1  & \textbf{3.62} & 4.10 & 3.35 & \textbf{9.94 G}  \\
    2 (For NR-LauraTTS)  & \textbf{3.62} & \textbf{4.11} & \textbf{3.36} & 9.96 G \\
    4  & 3.61 & 4.10 & 3.35 & 9.98 G   \\
    8  & 3.60 & 4.09 & 3.33 & 10.04 G  \\
    16 & 3.60 & 4.09 & 3.33 & 10.14 G  \\
    32 & 3.59 & 4.08 & 3.32 & 10.35 G  \\
    \bottomrule
  \end{tabular}}
\end{table}

\begin{table*}[t!]\scriptsize
  \caption{Comparison of subjective and objective evaluation results for the noise-robust zero-shot TTS task, where ``+FLOPs'' indicates the additional FLOPs compared to LauraTTS. Both MOS and SMOS results were reported with a 95\% confidence interval (CI).}
  \label{tab: tts}
  \centering
  \resizebox{0.9\textwidth}{!}{
  \begin{tabular}{c|ccccccc}
    \toprule
    Method                    & Prompt & MOS (CI) & SMOS (CI) & SECS & WER (\%) & CER (\%) & + FLOPs \\
    \midrule
    Ground Truth              & -      & 4.09 ($\pm$0.07) & - & 1.000 & 1.98 & 0.69  & - \\
    \midrule
    \multirow{2}{*}{LauraTTS \cite{du2023lauragpt}} & Clean  & 4.03 ($\pm$0.07) & 3.99 ($\pm$0.08) & 0.827 & 2.33 & 1.11  & -       \\
    & Noisy  & 3.46 ($\pm$0.09) & 2.80 ($\pm$0.11) & 0.642 & 21.37 & 17.31 & -       \\
    \midrule
    LauraTTS + MP-SENet\cite{lu2023mp} & \multirow{2}{*}{Noisy}  & 4.01 ($\pm$0.07) & 3.85 ($\pm$0.08) & 0.811 & 2.54 & 1.41  & 38.93 G \\
    \textbf{NR-LauraTTS} & & 4.02 ($\pm$0.07) & \textbf{3.98 ($\pm$0.07)}  & \textbf{0.827} & \textbf{2.44} & \textbf{1.27}  & \textbf{1.10 G} \\
    \bottomrule
  \end{tabular}}
\end{table*}

\subsection{Baselines and Evaluation Metrics}
We first conducted objective evaluations on the SE task to verify the effectiveness of discrete acoustic token denoising.
The proposed codec denoiser is compared with two SOTA SE methods, i.e., CMGAN \cite{cao2022cmgan} and MP-SENet \cite{lu2023mp}.
We followed the official implementations of CMGAN \footnote{\href{https://github.com/ruizhecao96/CMGAN}{https://github.com/ruizhecao96/CMGAN}.} and MP-SENet \footnote{\href{https://github.com/yxlu-0102/MP-SENet}{https://github.com/yxlu-0102/MP-SENet}.} to train the models using the constructed noisy-clean dataset.
The DNSMOS P.835 \cite{reddy2022dnsmos} model was adopted to evaluate the speech quality (SIG), background noise quality (BAK), and overall quality (OVRL) of the enhanced speech.
All DNSMOS metrics range from 1 to 5, with higher values indicating better performance.
Additionally, we used floating point operations (FLOPs) to assess the computational complexity of the models.
FLOPs were computed using 1-second speech signals as input to the models.

We then conducted objective and subjective evaluations to compare the noise-robust zero-shot TTS performance of our proposed NR-LauraTTS with the original LauraTTS and LauraTTS using the enhanced audio prompts by MP-SENet.
We used the audio samples from the test set with durations between 3 and 10 seconds.
For each sample synthesis, we randomly selected another utterance of the same speaker to serve as the audio prompt.
For objective evaluation, we assessed speaker similarity using the speaker encoder cosine similarity (SECS) from the Resemblyzer package \cite{louppe2019resemblyzer}. 
Additionally, we evaluated speech intelligibility using the word error rate (WER) and character error rate (CER), calculated with the Whisper \cite{radford2023robust} Large-V3 model to transcribe the synthesized speech.
For subjective evaluation, we conducted the mean opinion score (MOS) test and the similarity MOS (SMOS) test to evaluate the naturalness and speaker similarity of the synthesized speech.
We crowd-sourced over 30 raters on Amazon Mechanical Turk, with each rater scoring 20 speech samples on a scale from 1 to 5 in 0.5-point increments.

\section{Results and Analysis}
\subsection{Results on the SE Task}
The experimental results of the SE task are depicted in Table~\ref{tab: se}.
Overall, the proposed codec denoiser outperformed the SOTA SE methods CMGAN and MP-SENet among all the DNSMOS metrics, demonstrating that the speech enhanced by the proposed codec denoiser surpassed those of CMGAN and MP-SENet in speech quality, background noise quality, and overall quality.
Compared to CMGAN and MP-SENet, which conducted SE at the signal domain, our proposed codec denoiser operated in the discrete acoustic token domain, offering certain advantages in SE performance and reducing the model's computational complexity by at least threefold.
The performance superiority may be attributed to the fact that the VQs of the codec act as information bottlenecks, allowing the quantized discrete acoustic tokens to filter out some noise. 
Moreover, the first two groups of clean acoustic tokens contain more essential acoustic information, making them easier to predict.

We further conducted analysis experiments to evaluate the impact of varying the number of predicted acoustic token groups by the token denoiser on overall SE performance, with results shown in Table~\ref{tab: ablation}. 
As the number of predicted token groups increased, the computational complexity of the codec denoiser grew progressively, while the DNSMOS metrics initially improved until two predicted token groups were reached, after which they steadily declined. This verified that the first two groups of acoustic tokens contain sufficient acoustic information, and the greater the number of predicted token groups, the more challenging the prediction became, leading to degradation in the prediction accuracy.
Therefore, enhancing only the first two groups of acoustic tokens can both ensure the SE performance and reduce the model's complexity.

\subsection{Results on the Noise-Robust Zero-Shot TTS Task}
The experimental results of the noise-robust zero-shot TTS task are depicted in Table~\ref{tab: tts}.
Firstly, for the original LauraTTS, both subjective and objective metrics collapsed when using noisy audio prompts compared to the clean ones, highlighting the lack of noise robustness in LLM-based TTS methods, with noise in the audio prompt severely affecting the naturalness, intelligibility, and speaker similarity of the synthesized speech.
When using the SOTA SE model MP-SENet to enhance the noisy audio prompt before feeding it into LauraTTS, all metrics improved substantially, especially those related to speech naturalness and intelligibility. 
However, a gap remained in speaker similarity when compared to the LauraTTS model using clean audio prompts.
Finally, our proposed NR-LauraTTS, which enhanced the noisy audio prompt in the discrete acoustic token domain, achieved comparable performance to the LauraTTS with clean audio prompts among all the metrics, validating the accuracy of the token denoiser in predicting the first two groups of clean acoustic tokens and the effectiveness of our approach.

Compared to the approach that used MP-SENet to enhance the noisy audio prompts, our proposed NR-LauraTTS achieved higher speaker similarity, indicating that the artifacts introduced by SE models at the signal level would affect the speaker information in the enhanced audio prompts.
However, NR-LauraTTS enhanced the noisy audio prompts in the discrete acoustic token level by predicting the first two groups of clean acoustic tokens, which were considered to contain the primary acoustic properties like speaker identity, leading to a more accurate restoration of the speaker information and resulting in higher speaker similarity in the synthesized speech.
Additionally, since discrete acoustic tokens have lower temporal resolution compared to speech signals, our approach introduced only a minimal increase in the model's computational complexity compared to using an additional SE model, making it more suitable for practical applications.

\section{Conclusion}
In this paper, we proposed a codec denoiser for high-quality SE in the discrete acoustic token domain, and integrated it into the LauraTTS model to construct a noise-robust zero-shot TTS system, NR-LauraTTS.
The codec denoiser predicted the first two groups of enhanced acoustic tokens through a token denoiser, and used an embedding refiner to estimate the summed enhanced acoustic embedding, and finally reconstructed the enhanced speech waveform using the codec decoder.
Meanwhile, the enhanced acoustic tokens can be used as the acoustic prompt to the LauraTTS to produce high-quality personalized speech.
Experimental results demonstrated that the proposed codec denoiser outperformed the SOTA SE methods in enhanced speech quality, and the proposed NR-LauraTTS excelled in the approach using enhanced audio prompts with minimal impact on the model's computational complexity.
In future work, we will further investigate to achieve the controllability of acoustic environments in LLM-based zero-shot TTS models.

\section{Acknowledgements}
This work was funded by the National Nature Science Foundation of China under Grant U23B2053 and 62301521, and the Anhui Provincial Natural Science Foundation under Grant 2308085QF200.

\bibliographystyle{IEEEtran}
\bibliography{mybib}

\end{document}